\documentclass[12pt]{iopart}

\usepackage{dcolumn}
\usepackage{xcolor}
\usepackage[table]{colortbl}
\usepackage{multirow}
\usepackage{bm}
\usepackage{array,hhline} 
\usepackage{amssymb}
\usepackage{graphicx}
\usepackage{multicol}

\definecolor{Silver}{rgb}{0.9,0.9,0.9}

\begin{document}

\title[Orthogonal and covariant Lyapunov vectors]{Symmetry properties of orthogonal and  covariant Lyapunov vectors and their exponents}

\author{Harald A. Posch}

\address{Computational Physics Group, Faculty of Physics, University of Vienna}
\ead{harald.posch@univie.ac.at}

\date{\today}

\begin{abstract}
     Lyapunov exponents are indicators for the chaotic properties of a
classical  dynamical system. They are most naturally defined in terms of the time evolution of  a
set of so-called covariant vectors,  co-moving with the linearized flow in tangent space. 
Taking a simple spring pendulum and the H\'enon-Heiles system as examples, we demonstrate
the consequences of symplectic symmetry and of time-reversal invariance for such vectors, 
and study the transformation between different parameterizations of the flow. 
\end{abstract}

\pacs{05.45.-a,05.40.-a,05.20.-y,05.45.Pq}
\maketitle
%------------------------------------------------------------------------------
\section{Introduction}
       The stability of the phase-space trajectory of a dynamical system is determined by the so-called
Lyapunov exponents, which are the time-averaged rate constants of a set of perturbation vectors
in tangent space, which grow or shrink exponentially with time. Various such sets  have been 
introduced in connection with the algorithms  for the computation of the Lyapunov  exponents. The features
of these sets and of their associated exponents are well known to mathematicians and theoretical physicists
\cite{Oseledec:1968,Ruelle:1979,Eckmann:1985,Yang_Radons_2010}. In view of an increasing  
number of applications to ever more sophisticated  physical systems, it seems worthwhile to become familiar with  
the properties of these tangent-space objects for some simple models. This is the aim of the present paper. 
  
    The  most familiar set of perturbation vectors spanning the tangent space is the set of orthonormal Gram Schmidt (GS) vectors
$\{ ^{(+)}\bm{g}^{\ell}\} ,\; \ell \in \{1,\dots,D\}$,  commonly also  referred to as  {\em backward} Lyapunov vectors  
(since they depend on the history in the past). $D$ is the dimension of phase space.
For time-continuous systems they  are most elegantly obtained as the forward-in-time solution  (indicated here by an 
index (+)) of the linearized motion equations, augmented by a 
set of constraints, which continuously enforce the orthogonality and the norm conservation of these vectors. The latter  
gives rise to the GS-Lyapunov exponents along the way \cite{PH88,Goldhirsch}. Instead of constraints, the standard algorithms 
for the computation of Lyapunov exponents \cite{Benettin,Shimada,Wolf} employ a periodic  Gram-Schmidt re-orthogonalization
scheme, which may also be easily adapted for many-dimensional systems involving time-discontinuous maps 
such as the dynamics of hard spheres \cite{DPH}. 
 
    A second and less familiar set of  perturbation vectors are the covariant vectors $\{^{(+)}\bm{ v}^{\ell}\} ,\; \ell \in \{1,\dots,D\}$, 
 which are of unit length but generally not
orthogonal to each other. They evolve in tangent space according to the non-constrained
linearized motion equations. Still required is a periodic re-normalization, which also generates the corresponding
Lyapunov exponents. These vectors constitute a practical realization of the Oseledec  splitting
\cite{Oseledec:1968,Ruelle:1979,Eckmann:1985} of the tangent space into a hierarchy of stable and unstable subspaces 
at any phase-space point visited by the trajectory.

   In addition to these sets connected with the tangent flow forward in time, there exist 
corresponding sets, if the dynamics  is followed backward in time.  They will be  distinguished by  the index (-). Their application gives 
rise to the orthonormal {\em forward} Gram-Schmidt Lyapunov vectors  $\{ ^{(-)}\bm{g}^{\ell}\}$, which are conventionally  
called ''forward'' since they depend on their history in the future.
In general, they are not simply related to the set of backward Gram-Schmidt vectors.   Similarly, there exists a
set of time-reversed covariant vectors  $\{^{(-)}\bm{v}^{\ell}\}$, which, however,  for time-reversal invariant dynamics
agrees with its time-forward counterpart up to a simple reversal of the indices, $\ell \to D + 1 - \ell$.
   
    While the Gram-Schmidt vectors have been central to any algorithm since the pioneering days of the
numerical stability analysis  of dynamical systems  \cite{Benettin,Shimada,Wolf}  more than thirty years ago, 
the covariant vectors proved rather elusive due to the 
Lyapunov instability of the  computational process itself.  Practical schemes for the computation 
of covariant vectors have been developed only recently \cite{Ginelli,Wolfe}. A few studies of their properties for 
various systems have appeared since then \cite{Yang_Radons_2010,YRa,YRb,BP10,BPDH,PB,HH,Radons_2011,Kuptsov_2011}.  
In the following sections we shall study these properties for two simple symplectic systems, 
the chaotic spring pendulum  and the H\'enon-Heiles system. To establish our notation, we first summarize the most important relations
below.

\section{Definitions and notation}

If ${\bf \Gamma}(t)$ denotes the state of a dynamical system of dimension $D$,
its evolution equation and formal solution are given by
\begin{equation}
    \dot{\bf \Gamma} = {\bf F}({\bf \Gamma}),\;\;\;   {\bf \Gamma}(t) =  \phi^t ({\bf \Gamma}(0))
   \label{motion} 
\end{equation}     
where ${\bf F}$ is a (generally nonlinear) vector-valued function of dimension $D$, and 
the map $\phi^t$ defines the phase flow.  The linearized evolution equation and the formal solution
for an arbitrary perturbation vector $\delta {\bf \Gamma}(t)$ in tangent space become
\begin{equation}
    \dot{\delta {\bf \Gamma}} =  {\cal J}({\bf \Gamma})  \delta {\bf \Gamma} , \;\;\;\delta{\bf \Gamma}(t) =  D\phi^t \vert_{{\bf \Gamma}(0)}\;
\delta{\bf \Gamma}(0)
 \label{linearized}
\end{equation}    
where 
${\cal J}({\bf \Gamma}) =  \partial {\bf F} /  \partial {\bf  \Gamma}$, and  where $D \phi^t \vert_{{\bf  \Gamma}(0)}$ denotes the flow in tangent space.
As already mentioned, the covariant vectors evolve - co-rotate in particular - according to the unconstrained tangent-space flow,
\begin{equation}
\bm{v}^{\ell}\left( {\bf \Gamma}(t)\right)  = \frac{D \phi^t  \vert_{{\bf  \Gamma}(0)} \; \bm{v}^{\ell}\left({\bf \Gamma}(0)\right)}
                { \big\| D \phi^t  \vert_{{\bf  \Gamma}(0)} \; \bm{v}^{\ell}\left({\bf \Gamma}(0)\right)  \big\|}.
  \label{ev}
\end{equation}
  Of course, the algorithm has to ascertain that
the initial vector $ \bm{v}^{\ell}\left({\bf \Gamma}(0)\right) $     is already properly oriented and normalized in order to qualify as covariant.
The stretching  factor in the denominator of Equ. (\ref{ev}) provides the general definition for the (global)  Lyapunov exponents:
 \begin{equation}
^{(\pm)} \bar{\lambda}_{\ell}   =  \lim_{t \rightarrow \pm \infty}  \frac{1}{|t|}  \ln  \big\|  
 D\phi^t  \vert_{{\bf  \Gamma}(0)} \bm{v}^{\ell}\left({\bf \Gamma}(0)\right) \, \big\|, \;\;\; \ell \in \{1,\cdots,D\}
 \label{lyadef}
\end{equation}
Under very mild conditions, the multiplicative ergodic theorem of Oseledec \cite{Oseledec:1968,Eckmann:1985} asserts that the  global 
Lyapunov exponents are dynamical invariants and do not depend on the norm  and, hence,  the particular 
parameterization in  phase space. 

Two other limits of Eq. (\ref{lyadef}) are of special interest:
\begin{itemize}
\item $\tau$ large but finite:
 \begin{equation}
  \Lambda_{\ell}^{\makebox{cov}}(\tau)   =  \lim_{t \rightarrow \tau < \infty}  \frac{1}{|t|}  \ln  \big\|  
 D\phi^t  \vert_{{\bf  \Gamma}(0)} \bm{v}^{\ell}\left({\bf \Gamma}(0)\right) \, \big\|,  \; \;  \tau  \;\; \makebox{large but finite}.
 \label{ftle}
\end{equation}
These objects  depend on $\tau$ and are referred to as 
finite-time Lyapunov  exponents (FTLE). For infinite time they converge to  the global exponents $\bar{\lambda}_{\ell}$.
For finite $\tau$ they are no dynamical invariants and depend, for example, on the coordinate system in use. 
However, it was  recently shown by Kuptsov and Politi \cite{Kuptsov_2011} that the fluctuations of these quantities, i.e. the
linear  growth rate of the covariances of the logarithmic expansion factors $\tau \Lambda_{\ell}^{\mbox{cov}}(\tau) $,
\begin{equation}
   D_{\ell\ell'}^{\mbox{cov}}(\tau)  = \lim_{\tau < \infty } \left[ \left\langle \Lambda_{\ell}^{\makebox{cov}}(\tau) \Lambda_{\ell'}^{\makebox{cov}}(\tau) \right\rangle  - 
           \bar{\lambda}_{\ell} \bar{\lambda}_{\ell'}  \right]   \tau,
      \label{diff_matrix}     
\end{equation}   
is a dynamical invariant for large-enough $\tau$. Here, $\langle \dots \rangle$ denotes an average over (infinitely) many uncorrelated realizations
along a trajectory. Below we shall demonstrate this property for the chaotic pendulum.
\item $\tau \to 0$:
The limit
 \begin{equation}
  \Lambda_{\ell}^{\makebox{cov}}({\bf \Gamma}(0))   =  \lim_{\tau \rightarrow 0}  \frac{1}{|\tau|}  \ln  \big\|  
 D\phi^{\tau}  \vert_{{\bf  \Gamma}(0)} \bm{v}^{\ell}\left({\bf \Gamma}(0)\right) \, \big\|, \;\;\; \ell \in \{1,\dots,D\}
\label{D_covariance} 
\end{equation}
provides a definition of the so-called {\em local} Lyapunov exponents (LLE). They are point functions in phase space. 
Their time average over a long trajectory also converges to the
global exponents $\bar{\lambda}_{\ell}$.  If the covariant vector $\bm{v}^{\ell}({\bf \Gamma})$ is known at a point ${\bf \Gamma}$
the corresponding covariant LLE follows from
\begin{equation}
    \Lambda_{\ell}^{\mbox{cov}}({\bf \Gamma}) = 
    \left[\bm{v}^{\ell}({\bf \Gamma})\right]^T   {\cal J}({\bf \Gamma}) \; \bm{v}^{\ell}({\bf \Gamma}),
 \label{con}
 \end{equation}   
 where $T$ means transposition and ${\cal J}$ is the Jacobian of Eq. (\ref{linearized}). This nicely underlines the local nature of the LLEs,
 which depend implicitly on time.

\end{itemize}

So far all definitions of Lyapunov exponents are in terms of  covariant  Lyapunov vectors.   There are  
 analogous definitions for the orthonormal Gram-Schmidt vectors both forward and backward in time,  which are summarized in 
 Ref. \cite{BPDH} and are not repeated here. The indices GS and cov will be used in the following to distinguish between the quantities. 
  The GS-FTLEs also converge to the global exponents with $\tau \to \infty$, 
 as do the GS-LLEs when time-averaged along a trajectory. It is interesting to note that the 
 algorithm with continuously constrained orthonormality mentioned above  \cite{PH88,Goldhirsch,BPDH}
 provides an expression for the GS-LLEs,
  \begin{equation}
    \Lambda_{\ell}^{\mbox{GS}}({\bf \Gamma}) = 
    \left[\bm{g}^{\ell}({\bf \Gamma})\right]^T   {\cal J}({\bf \Gamma}) \;  {\bf g}^{\ell}({\bf \Gamma}),
 \label{continb}
 \end{equation}   
 which is analogous to Eq. (\ref{con}) for the covariant case.

The classical algorithms involving GS re-orthonormalization keep track of the volume changes 
of $d$-dimensional volume elements in phase space ($d \le D$).
For symplectic systems, for which the phase volume is invariant, the GS-LLEs show 
 the symplectic local pairing symmetry  \cite{Ramaswamy,BPDH,PB},    
\begin{eqnarray}
^{(+)}\Lambda_{\ell}^{\mbox{GS}}(t) &=&  -^{(+)}\Lambda_{D + 1 -\ell }^{\mbox{GS}}(t) \label{pair1}\\
^{(-)}\Lambda_{\ell}^{\mbox{GS}}(t) &= &  -^{(-)}\Lambda_{D + 1 -\ell }^{\mbox{GS}}(t) , \label{pair2}
\end{eqnarray}
where $^{(+)}$ and $^{(-)}$ indicate  whether the trajectory is followed forward or
backward in time. Since the angle information is discarded by the GS-process, these exponents
do not show time-reversal symmetry,
\begin{equation}
^{(-)}\Lambda_{\ell}^{\mbox{GS}}(t) \ne -^{(+)}\Lambda_{D + 1 -\ell }^{\mbox{GS}}(t). \label{inequ}
\end{equation}
 If the system is not symplectic, also Eqs. (\ref{pair1}) and  (\ref{pair2}) cease to exist.
On the other hand,  the orientation of the covariant 
vectors is not affected by any process such as re-orthogonalization, and  the covariant LLEs display  the time reversal
symmetry \cite{Ruelle:1979,BPDH}:
\begin{equation}   
 ^{(-)}\Lambda_{\ell}^{\mbox{cov}}(\Gamma(t))  = 
 - ^{(+)}\Lambda_{D+1 -\ell}^{\mbox{cov}}(\Gamma(t)) \; ;  \;\;\;  \ell = 1,\cdots,D.
 \label{local_symmetry_e}
\end{equation}
Similarly, the covariant vectors obey
\begin{equation}   
 ^{(-)}\bm{v}^{\ell}(\Gamma(t))  = 
 \pm  ^{(+)}\bm{v}^{D+1 -\ell}(\Gamma(t)). 
 \label{local_symmetry_v}
\end{equation}
These relations apply whether or not the system is symplectic, as long as it is time reversible.

   Another attractive property of the covariant vectors derives from the fact that they are
spanning sets for the stable and unstable Oseledec subspaces associated with the respective
Lyapunov exponents  (one-dimensional without degeneracy, and $m$-dimensional for multiplicity $m$),
and are defined without reference  to a particular parameterization.  This does not apply to the 
GS-exponents which suffer from the additional constraint of orthogonality. It has been shown recently
by Yang and Radons that covariant vectors  may be easily transformed
from one coordinate system to another, and the same is true for the  covariant finite-time and local Lyapunov exponents
\cite{Yang_Radons_2010} We shall demonstrate this property for the chaotic pendulum below by transforming from the
Cartesian representation to a polar coordinate system.

In the following  section we consider the chaotic spring pendulum, and in Section 3  we discuss
 the H\'enon-Heiles system.  We close with a short survey of the merits and disadvantages
 of the covariant analysis.

\section{The spring pendulum}

We consider  the planar chaotic motion of the spring pendulum \cite{PHH,DH} as a simple example. 
It consists of a mass $m$ attached to a harmonic spring with spring constant $k$ and rest length $R$, which 
is exposed to a
homogeneous gravitational force of strength $m g$ in the negative $y$ direction. The spring is attached to a pivot at the origin. The Hamiltonian in Cartesian coordinates $(x,y)$ and with the conjugate momenta 
$(p_x,p_y) = m(\dot{x}, \dot{y})$ is given by
\begin{equation}
   H_C = \frac{p_x^2 + p_y^2}{2m} + \frac{k}{2} \left( \sqrt{x^2 + y^2} - R \right)^2 + mgy  \label{cartesian},
\end{equation}
from which follow the equations of motion for the reference trajectory in phase space,
\begin{equation}
\begin{array}{rcl}
\dot{x} &=& p_x/m   \\
\dot{y} &=& p_y/m  \\
\dot{p}_x &=& k [ (R/r) -1] x   \\
\dot{p}_y &=& k [ (R/r) -1] y  - m g, 
\end{array}   
 \label{ceom}
\end{equation}
and for the perturbation vectors in tangent space.
\begin{equation}
\begin{array}{rcl}
\dot{\delta x} &=& \delta p_x / m\\
\dot{\delta y} &=& \delta p_y / m\\
\dot{\delta p_x} &=& k [(R/r) - 1] \delta x - (k R x/ r^3) (x \delta x + y \delta y) \\
\dot{\delta p_y} &=& k [(R/r) - 1] \delta y - (k R y/ r^3) (x \delta x + y \delta y),
\end{array}
 \label{peom}
\end{equation}
Here, $r = \sqrt{x^2 + y^2}.$

     Introducing polar coordinates $(r,\phi)$ through
\begin{equation}
     x = r \sin \phi, \;\; \;\;
     y = r \cos \phi , \label{cp}
\end{equation} 
the Hamiltonian becomes
\begin{equation}
    H_P = \frac{p_r^2}{2m} + \frac{p_{\phi}^2}{2m r^2} + \frac{k}{2} (r - R)^2 + m g r \cos \phi,
\end{equation}    
where $p_r = m \dot{r} $ and $p_{\phi} = m r^2 \dot{\phi}$ are the conjugate momenta.
The equations of motion in phase space now read
\begin{equation}
\begin{array}{rcl}
\dot{r} &=& p_r/m \\
\dot{\phi} &=& p_{\phi} /(m r^2) \\
\dot{p_r} &=& p_{\phi}^2 / (m r^3) - k (r - R) - m g \cos \phi \\
\dot{p_{\phi}} &=& m g r \sin \phi, 
\end{array}
 \label{p}
\end{equation}
from which the linearized motion equations for the perturbation vectors readily follow,
\begin{equation}
\begin{array}{rcl}
\dot{\delta r} &=& \delta p_r / m \\
\dot{\delta \phi} &=& \delta p_{\phi} / (m r^2) - 2 p_{\phi} \delta r /(m r^3)  \\
\dot{\delta p_r} &=& 2 p_{\phi} \delta p_{\phi}/(m r^3) - 3 p_{\phi}^2 \delta r /(m r^4) - k \delta r 
                             +  m g \sin \phi \delta \phi \\
 \dot{\delta p_{\phi}} &=& m g \sin \phi \delta r + m g r \cos \phi \delta \phi.           
\end{array}
 \label{pol}
\end{equation}

Let us consider the state point ${\bf \Gamma}_C = (x,y, p_x,p_y)$, which in the polar coordinate system
becomes ${\bf \Gamma}_P = (r, \phi, p_r, p_{\phi})$. This, according to Eq. (\ref{cp}), is
accomplished by the transformation \cite{Yang_Radons_2010}
\begin{equation}
  {\bf \Gamma}_P \equiv {\cal P}( {\bf \Gamma}_C) =  \left( \sqrt{x^2+y^2}, \tan^{-1} \left(\frac{x}{y}\right), 
        \frac{x p_x + y p_y}{ \sqrt{x^2 + y^2}},  (p_x y - x p_y)\right)^T. \label{transf}
\end{equation}  
Any Cartesian covariant vector $\bm{v}_{C}^{\ell}$ is transformed to the polar representation
according to 
\begin{equation}
\delta{\bf \Gamma}_{P}^{\ell} = {\cal M}  \; \bm{v}_{C}^{\ell},
\label{repr}
\end{equation}
where ${\cal M} = \partial {\cal P} / \partial  {\bf \Gamma}_C$ is the Jacobian of the transformation (\ref{transf}).
A final normalization,
\begin{equation}
 \bm{v}_{P}^{\ell} = \pm \frac{\delta{\bf \Gamma}_{P}^{\ell}  }
   {\big\|\delta{\bf \Gamma}_{P}^{\ell}\big\|} = 
          \pm \frac{ {\cal M}   \; \bm{v}_{C}^{\ell}} { \big\| {\cal M}  \; \bm{v}_{C}^{\ell} \big\|},
 \label{vrel}         
 \end{equation}
yields the covariant  vector in the polar representation. Although formulated in terms of a
particular coordinate transformation, the expressions (\ref{repr}) and (\ref{vrel}) are
completely general \cite{Yang_Radons_2010}.

For our numerical work we use reduced units for which the mass $m$, the rest length $R$, 
and the gravitational acceleration $g$ are unity.  The spring constant $k$ is set to two.
For the initial condition of the reference trajectory we take $(x,y,p_x,p_y) = (0.00001,1,0,0)$,
which determines the energy.
Care must be taken that the reference trajectory in the Cartesian and polar representations
coincide. Therefore, only Eqs. (\ref{ceom}) are integrated, and the solutions of the polar
equations (\ref{p}) are obtained by the Cartesian-to-polar transformation (\ref{transf}).
The computation of the covariant vectors and their associated local Lyapunov exponents has
been outlined in Refs. \cite{Ginelli,BP10,BPDH}, to which we refer for details.
The integration is carried out with a 4th-order Runge-Kutta algorithm with a time step of 0.002.
The global Lyapunov spectrum is found to be $\{ 0.0565, 0, 0, -0.0565 \}$. 
%%%%%%%%%%%%%%%%%%%%%%%%%%%%%%%%%%%%%%%%%%%%
\begin{figure}[htb]
\center
\vspace{-3cm}
\includegraphics[width=0.8\textwidth]{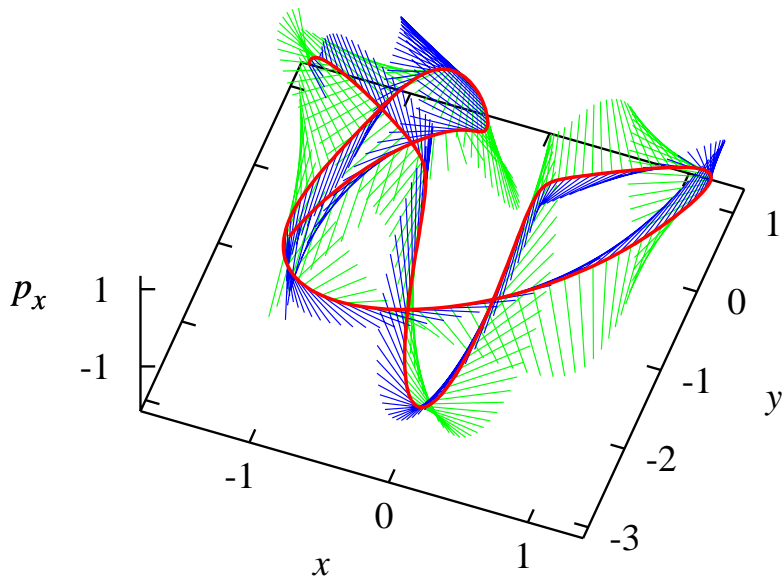}\\
\vspace{-1.0cm}
\caption{(Color online) Projection of a short spring-pendulum trajectory onto the 
$(x,y,p_x)$-subspace.  The blue and green lines represent projections of the covariant vectors
 $\bm{v}^1$ and $\bm{v}^4$, respectively, and give an impression of the respective 
 unstable and stable manifolds along the trajectory. The time interval for this trajectory segment 
 is 16 time units.} 
\label{figure_1}
\end{figure}
%%%%%%%%%%%%%%%%%%%%%%%%%%%%%%%%%%%%%%%%%%%%%%

By the smooth (red) line in Fig. \ref{figure_1} we show a projection of a short  trajectory into the 
three-dimensional subspace spanned by $x,y,p_x$. The covariant vector $\bm{v}^1$, which spans the
unstable manifold, is indicated in green. Similarly, the covariant vector $\bm{v}^4$ spanning the stable
manifold is shown in blue. Of course, these projected vectors are not of unit length.  The figure may provide
an intuitive understanding of the directions of maximum stretching or contraction of perturbations in
phase space.

      The time evolution of the polar components of  $\bm{v}^4$ is 
depicted in Fig. \ref{figure_2}. This vector was chosen for display since it is the most time consuming to compute
and has a non-vanishing global exponent.
The red lines are the results of a  direct integration of Eqs. ({\ref{pol}) within the polar
environment, whereas the green lines are obtained by converting the Cartesian covariant vector
$\bm{v}_C^4$ to its polar representation. As is indicated in Eq. (\ref{vrel}), the
sign is irrelevant since only the sense of direction counts. The same agreement is also obtained
for the other covariant vectors (not shown).
%%%%%%%%%%%%%%%%%%%%%%%%%%%%%%%%%%%%%%%%%%%%%%%%%%%
\begin{figure}[htb]
\vspace{-1cm}
\begin{tabular}{c c}
\\
\begin{minipage}[c]{.5\linewidth}
\includegraphics[width=1\textwidth]{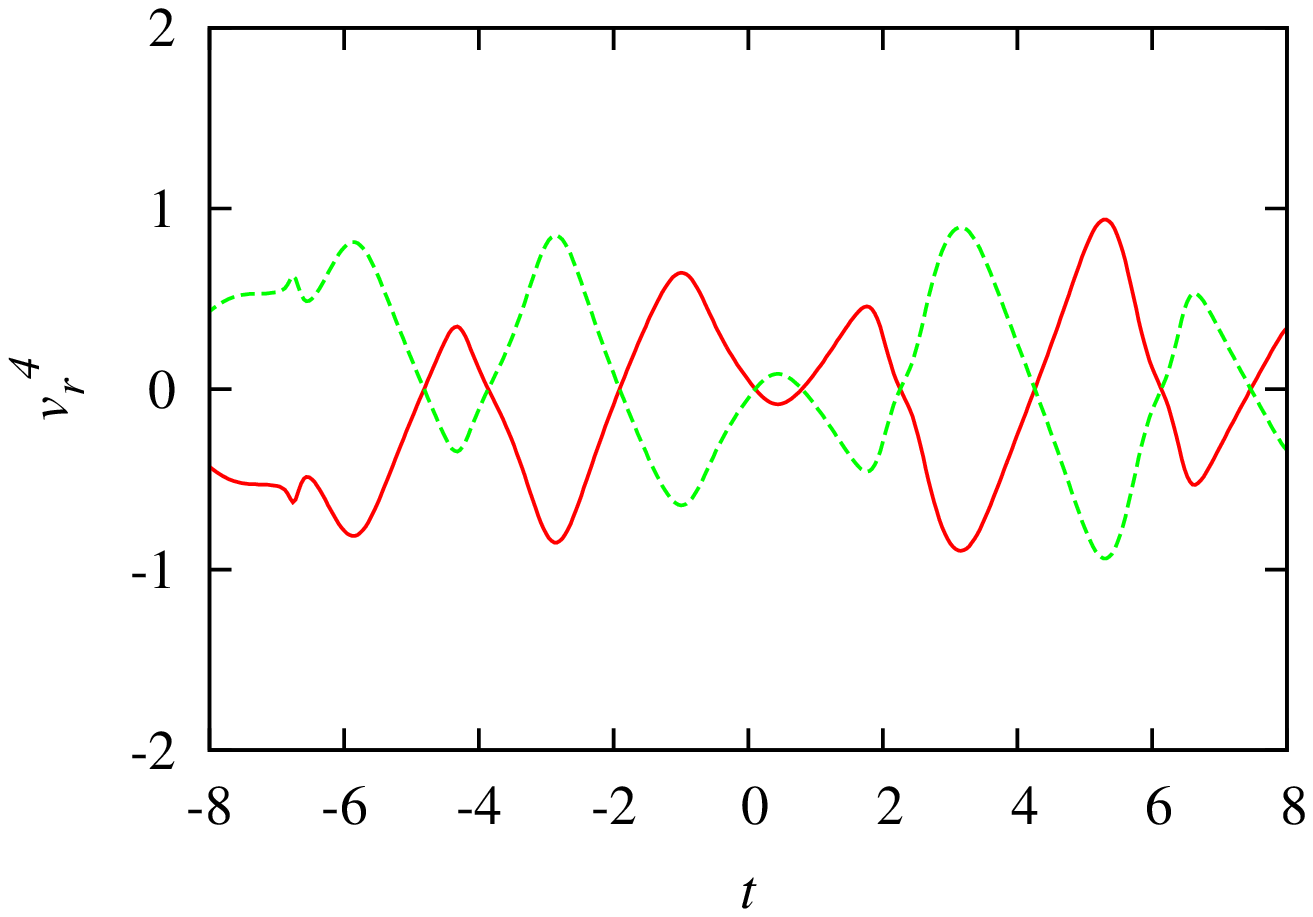}
\end{minipage} &
\begin{minipage}[c]{.5\linewidth}
\includegraphics[width=1\textwidth]{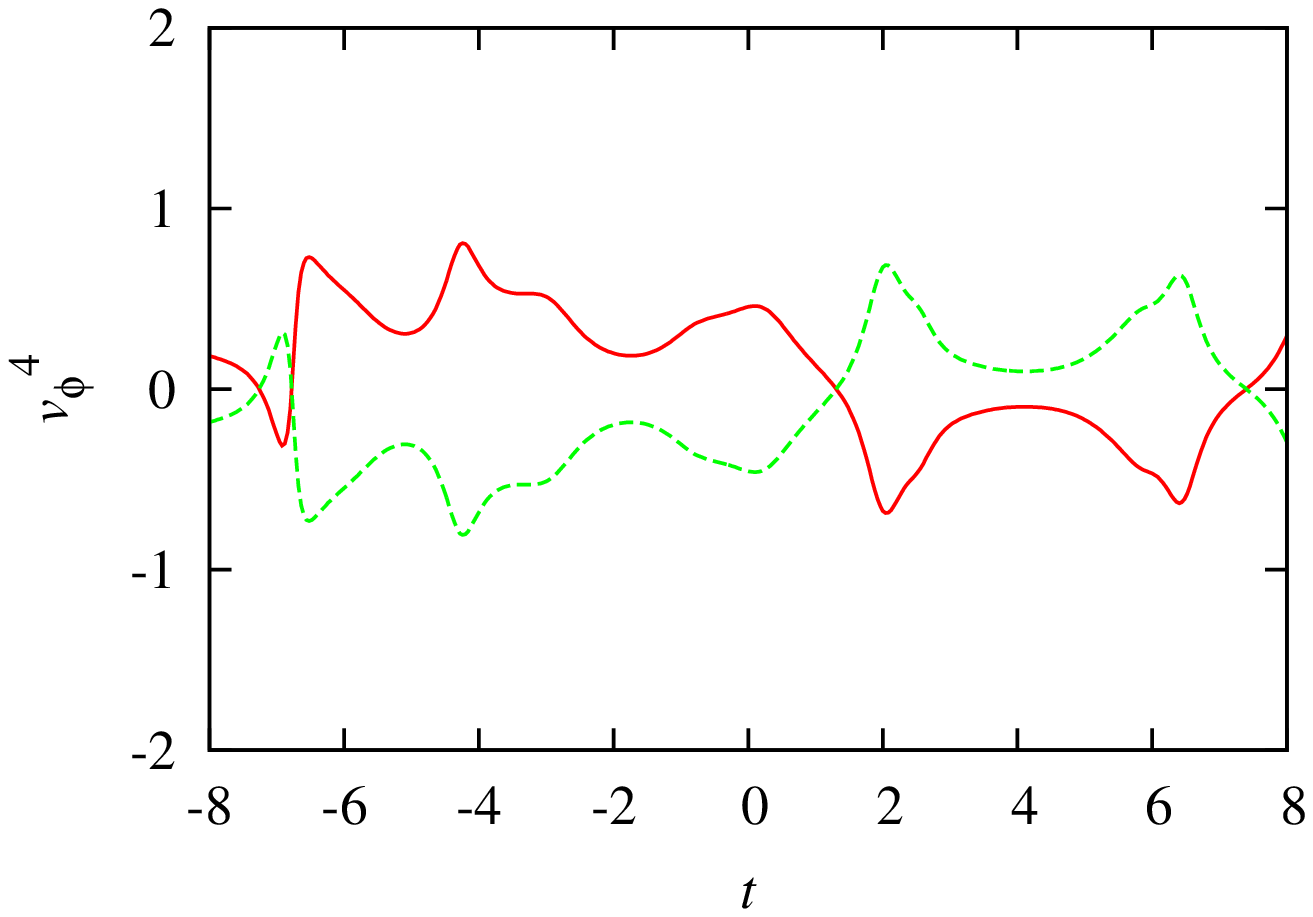}
\end{minipage} \\
\begin{minipage}[c]{.5\linewidth}
\includegraphics[width=1\textwidth]{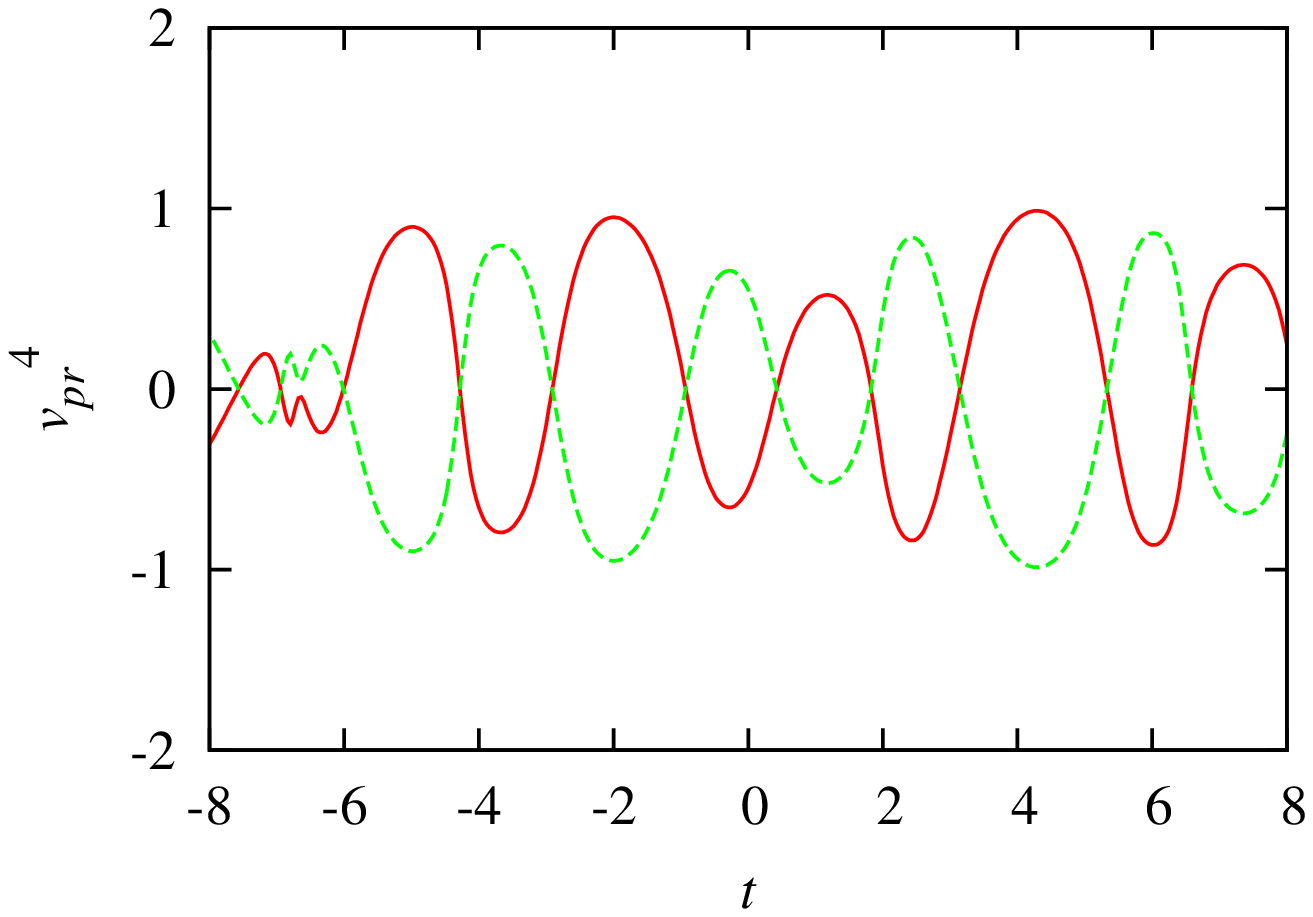}
\end{minipage}&
\begin{minipage}[c]{.5\linewidth}
\includegraphics[width=1\textwidth]{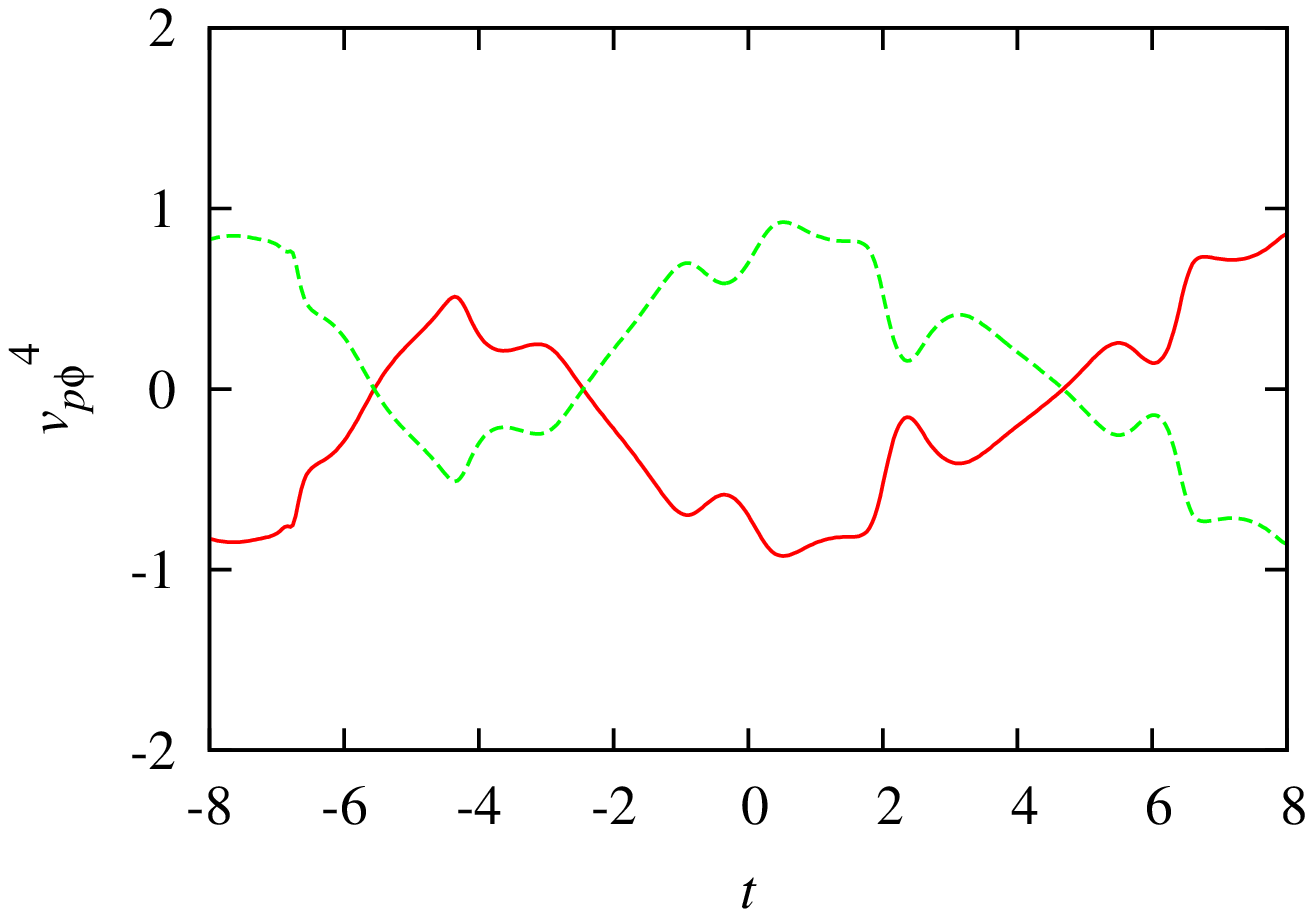}
\end{minipage}
\end{tabular}
\caption{(Color online) Time evolution along the trajectory segment depicted in Fig. \ref{figure_1} of the 
four polar components $\delta r, \delta{\phi}, \delta p_r, \delta p_{\phi}$ (from top left to bottom right) for  the covariant vector  $\bm{v}_P^4$ spanning the stable manifold of the spring pendulum. The smooth (red) lines
are obtained by direct integration with polar coordinates, whereas the dashed (green) curves were converted from the Cartesian representation.}
\label{figure_2}
\end{figure}
%%%%%%%%%%%%%%%%%%%%%%%%%%%%%%%%%%%%%%%%%%%%%%%%%%%

    The finite-time Lyapunov exponents (including the limiting case of local exponents) in the Cartesian and polar 
    representations are also intimately  connected. Let us consider a time element of duration $t_{n+1} - t_n = \tau$. The Cartesian
covariant vector $\bm{v}_C^{\ell}(t_n)$ evolves according to Eqs.  (\ref{ev}) and (\ref{ftle})  
into a vector $\exp \left\{\Lambda_C^{\ell,\mbox{cov}} \tau\right\} \bm{v}_C^{\ell}(t_{n+1})$, which --
according to Eq. (\ref{repr})  -- may be transformed to the polar representation to give
%%%%%%%%%%%%%%%%%%%%%%%%%%%%%%%%%%%%%%%%%%%%%%
\begin{figure}[htb]
\begin{tabular}{c c}
\\
\begin{minipage}[c]{.5\linewidth}
\includegraphics[width=1\textwidth]{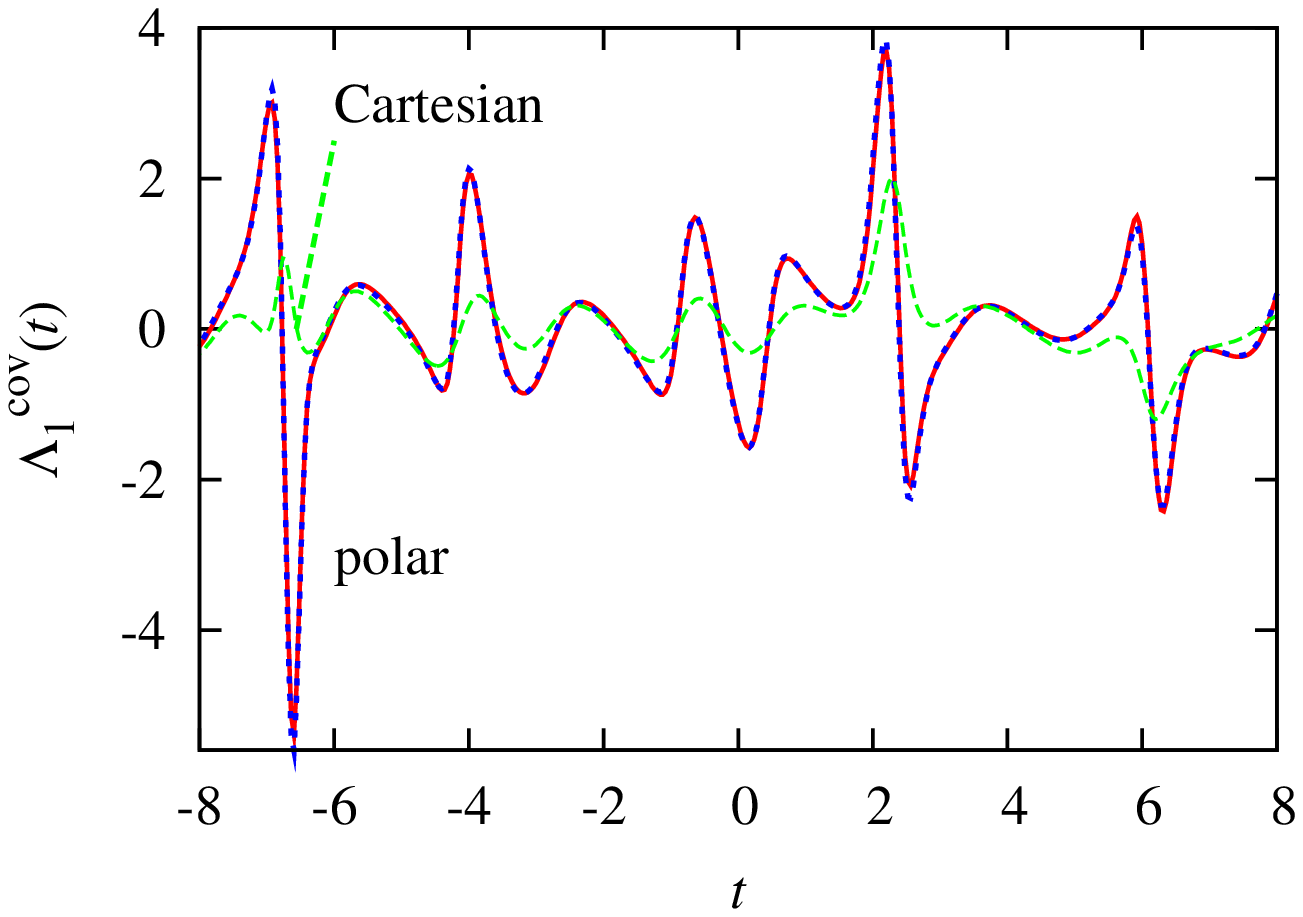}
\end{minipage} &
\begin{minipage}[c]{.5\linewidth}
\includegraphics[width=1\textwidth]{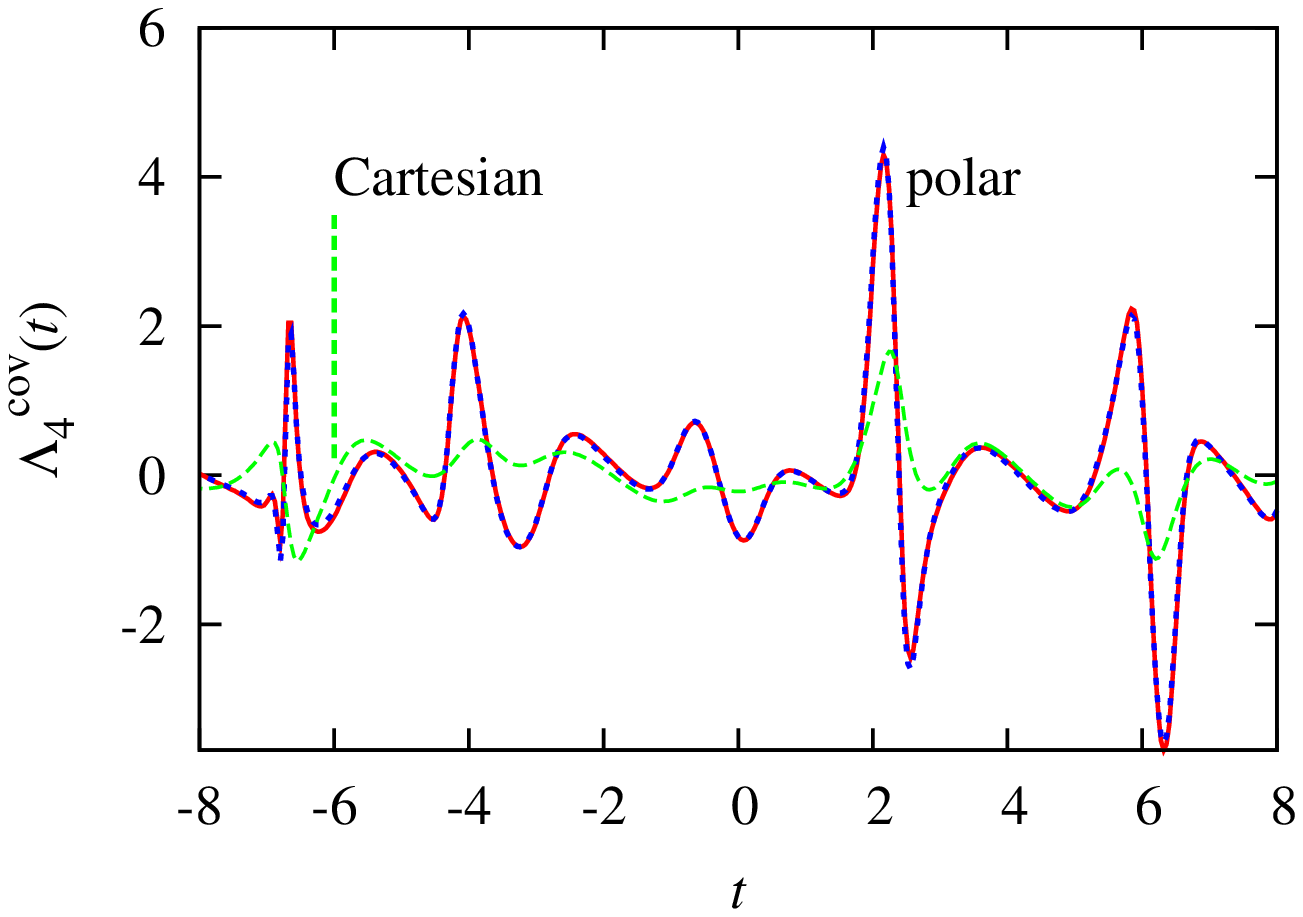}
\end{minipage} \\
\end{tabular}
\caption{(Color online) Comparison of time dependent local covariant exponents for
the unstable ($\ell = 1$, left panel) and stable ($\ell = 4$, right panel) manifolds of the 
spring pendulum. The smooth (red) lines are obtained by numerical integration of the 
polar equations of motion, the  dotted (blue) lines by integration of the Cartesian equations
of motion, followed by a conversion to the polar representation. The agreement is very good.
For comparison, the corresponding Cartesian exponents, which are the basis of the conversion,
 are shown by the dotted (green) lines.}
 \label{figure_2a}
\end{figure}
%%%%%%%%%%%%%%%%%%%%%%%%%%%%%%%%%%%%%%%%%%%%%%
$$\exp \left(\Lambda_C^{\ell, \mbox{cov}} \tau \right) {\cal M}(t_{n+1}) \bm{v}_C^{\ell}(t_{n+1}). $$ 
Alternatively,  $\bm{v}_C^{\ell}(t_n)$ at $t_n$ may be first transformed to the polar representation,
${\cal M}(t_n) \bm{v}_C^{\ell}(t_n)$, and then be evolved in 
time to the end of the interval, which yields the vector
$$ \exp \left(\Lambda_P^{\ell, \mbox{cov}} \tau \right) \big\| {\cal M}(t_n) \bm{v}_C^{\ell}(t_n) \big\|
\bm{v}_P^{\ell}(t_{n+1}). $$ Equating these two expressions and taking the absolute value
on both sides yields
\begin{equation}
\Lambda_P^{\ell, \mbox{cov}} = \Lambda_C^{\ell, \mbox{cov}} + 
  \frac{1}{\tau} \ln \frac{\big\| {\cal M}(t_{n+1})\: \bm{v}_C^{\ell}(t_{n+1}) \big\|}
                          {\big\| {\cal M}(t_{n}) \; \bm{v}_C^{\ell}(t_{n}) \big\|}.
                         \label{relate_exponents}                   
 \end{equation}
This provides the relation between the covariant local exponents in the two coordinate systems. 
Although expressed here in terms of the Cartesian and polar coordinates, the
expressions (\ref{vrel}) and (\ref{relate_exponents}) are completely general.

Eq. (\ref{relate_exponents}) clearly shows that the local exponents differ for different coordinate systems 
\cite{Yang_Radons_2010}.
To test the last relation, we show by the smooth (red) lines in Fig. \ref{figure_2a} the local 
(time-dependent) covariant exponents in the polar representation for the 
unstable manifold ($\ell = 1$, left panel) and for the stable manifold ($\ell = 4$, right panel)
of the spring pendulum. They are obtained by direct integration
of the motion equations (\ref{pol}) with polar coordinates.  For comparison, the corresponding
Cartesian exponents are plotted by the dashed (green) lines. If the latter are transformed
to the polar representation, the dotted (blue) lines are obtained.
The perfect match with the smooth (red) lines of the strictly polar approach asserts the validity
of Eq. (\ref{relate_exponents}). 

  In Fig. \ref{figure4} we show the (linear) growth rate of the  covariance matrix for the logarithmic 
  expansion $\tau \Lambda_{\ell} (\tau)$ as a function of the averaging interval $\tau$ for  various representations.
  For the  GS-vectors this rate matrix is given by \cite{Kuptsov_2011}
\begin{equation}
   D_{\ell\ell'}^{\mbox{GS}}(\tau)  = \lim_{\tau < \infty } \left[ \left\langle \Lambda_{\ell}^{\makebox{GS}}(\tau) 
        \Lambda_{\ell'}^{\makebox{GS}}(\tau) \right\rangle  -   \bar{\lambda}_{\ell} \bar{\lambda}_{\ell'}  \right]   \tau,
      \label{GS_diff_matrix}     
\end{equation}   
where we still have to distinguish the cases with Cartesian or polar coordinates. For covariant vectors the 
definition is given by Eq. (\ref{diff_matrix}). For clarity  we restrict ourselves to matrix elements, which do not involve 
any of the vanishing Lyapunov exponents $\bar{\lambda}_2$ and $\bar{\lambda}_3$.  The panel on the right-hand side
is a magnification of the small-$\tau$ regime of the left panel. The elements for
Gram-Schmidt FTLEs with Cartesian coordinates are indicated by red full points (case GS-C),
those with polar coordinates by the blue full squares (case GS-P).  Similarly, the elements involving covariant FTLEs 
with Cartesian coordinates are indicated by green crosses (case COV-C), those with polar coordinates by  
violet open circles (case COV-P).  The following observations are made: 
\begin{enumerate}
\item There exists the general symmetry $D_{11}(\tau) = D_{44}(\tau)$ and $D_{14}(\tau) = D_{41}(\tau)$ for any coordinate system and
for any vector set used, be it Gram-Schmidt or covariant.  Wheras the latter equality is trivial, the former is not in view of
the fact that neither $\bm{v}^1$,$\bm{v}^4$ nor   $\bm{g}^1$,$\bm{g}^4$  are simply related to each other.
\item  $D_{11}(\tau)$ for GS-C and COV-C agree for all $\tau$, and similarly $D_{11}(\tau)$ for GS-P and COV-P.   
This is to be expected, since the covariant vector $\bm{v}_1$ and the GS vector ${\bm{g}_1}$ associated with the maximum exponent
are identical by construction. 
\item For small averaging intervals $\tau$, the fluctuations of the finite-time exponents as measured by $D_{11}(\tau)$ and $D_{14}(\tau)$  
significantly differ for the cases GS-C and GS-P, as has been observed before  \cite{PHH}.  A similar observation is made for the
cases COV-C and COV-P.  These differences gradually disappear for averaging times  $\tau > 10$. 
\item  The cross correlation $D_{14}(\tau)$ for the covariant and Gram-Schmidt vector sets differ enormously for small $\tau$,
regardless what coordinate system is used. The difference is even larger between COV-P and GS-P than between COV-C and GS-C.
By increasing $\tau$ these differences very slowly disappear.
\item From the left panel of Fig. \ref{figure4} one infers that for $\tau > 200$ all elements of the covariance matrix for the FTLEs 
become independent of the choice of the set of perturbation vectors and of the coordinate system. Thus, $\lim_{\tau \to \infty} D(\tau)$ 
becomes a dynamical invariant independent of the metric and parameterization \cite{Kuptsov_2011}.  For high-dimensional chaotic 
systems such a result  has been interpreted as
evidence for effective hyperbolicity of the dynamics  and of the statistical insignificance of hyperbolic tangencies. 
\end{enumerate}
%%%%%%%%%%%%%%%%%%%%%%%%%%%%%%%%%%%%%%%%%%%%%
\begin{figure}[htb]
\begin{tabular}{c c}
\\
\begin{minipage}[c]{.5\linewidth}
\includegraphics[width=1\textwidth]{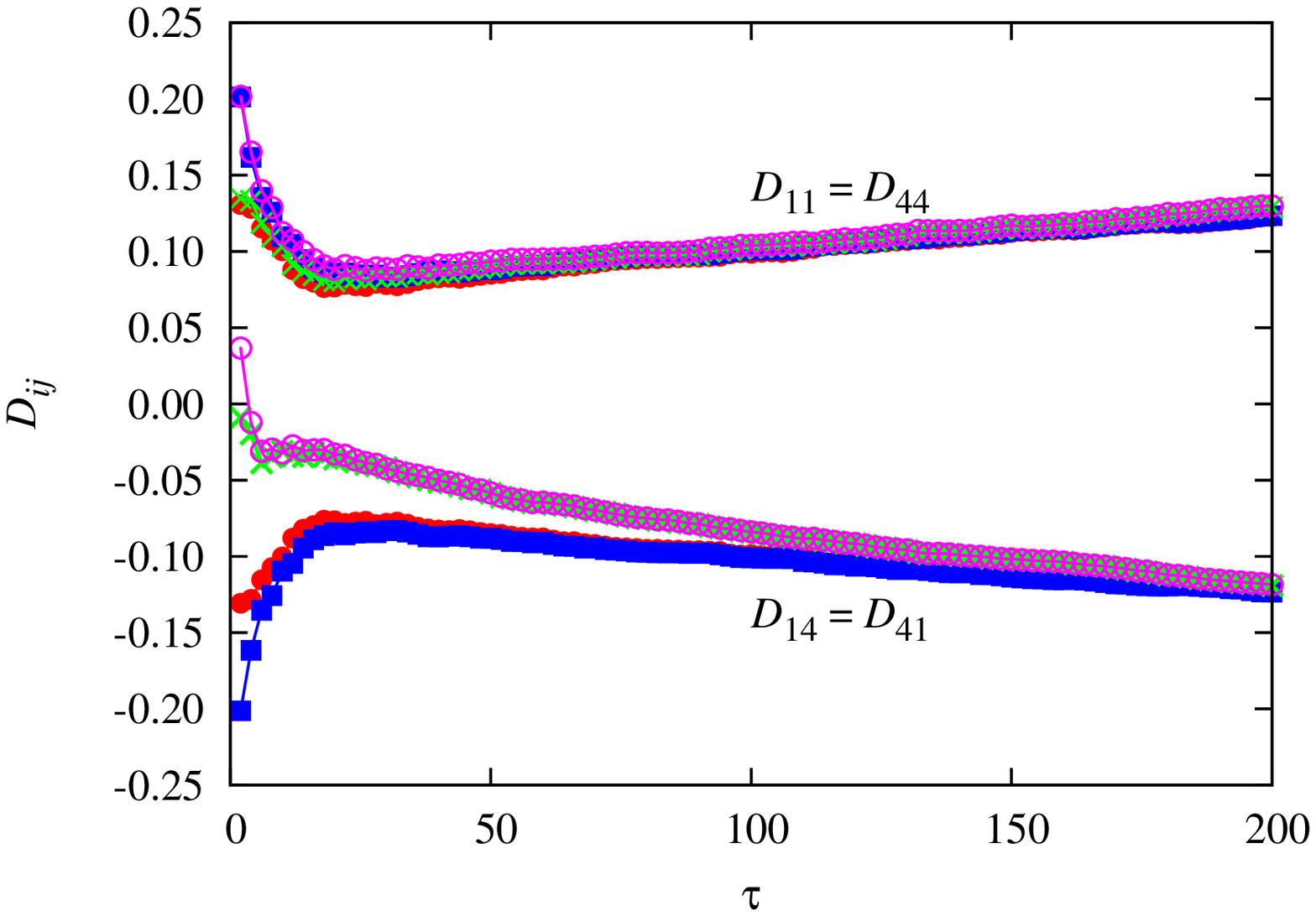} 
\end{minipage} &
\begin{minipage}[c]{.5\linewidth}
\includegraphics[width=1\textwidth]{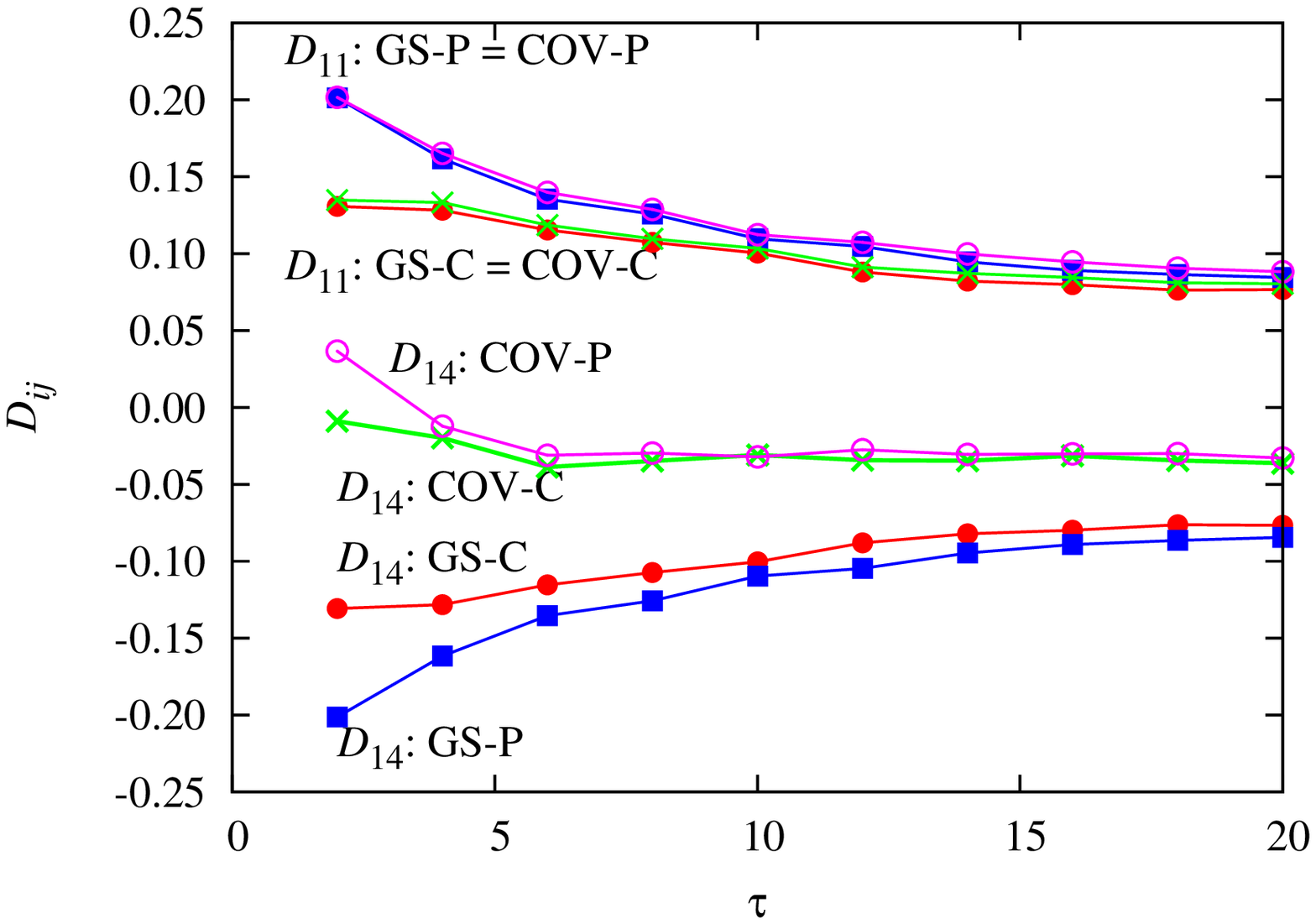}
\end{minipage} \\
\end{tabular}
\caption{(Color online) Chaotic pendulum: $\tau$ dependence of the
 growth rate of the  covariance matrix for the logarithmic expansion $\tau \Lambda_{\ell} (\tau)$. 
 Only matrix elements not involving vanishing global exponents are shown.
The panel on the right-hand side is a magnification of the small $\tau$ regime.  
Red full points: Cartesian coordinates and GS exponents (case GS-C); 
blue full squares: polar coordinates and GS exponents (case GS-P); 
green crosses: Cartesian coordinates and covariant exponents (case COV-C):
violet open circles: polar coordinates and covariant exponents (case COV-P).
In all cases the evolution is in the direction of the positive time axis.}
\label{figure4}
\end{figure}
%%%%%%%%%%%%%%%%%%%%%%%%%%%%%%%%%%%%%%%%%%%%%%

%%%%%%%%%%%%%%%%%%%%%%%%%%%%%%%%%%%%%%%%%%%%%%%%%

\section{The H\'enon-Heiles system}

As a second example, we consider
the familiar H\'enon-Heiles system \cite{Henon} with a Hamiltonian in Cartesian coordinates
\begin{equation}
 H_C= \frac{1}{2} (p_x^2 + p_y^2 ) + \frac{1}{2} ( x^2 + y^2 ) + x^2 y - \frac{1}{3} y^3.
 \label{henon}
\end{equation}
For an energy $1/6$, the system is known to be chaotic
(with a global Lyapunov spectrum $\{ 0.127_7, 0, 0, -0.127_7 \}$), where the trajectory visits 
most of the accessible phase space \cite{LL_Buch,Ramaswamy}. The local Lyapunov
exponents display the
symplectic pairing symmetry Eq. (\ref{pair1}-\ref{pair2}), and the time reversal symmetry 
of Eq. (\ref{local_symmetry_e}), as is demonstrated in the respective left and right panels
of Fig. \ref{figure_3}.

%%%%%%%%%%%%%%%%%%%%%%%%%%%%%%%%%%%%%%%%%%%%%%
\begin{figure}[htb]
\begin{tabular}{c c}
\\
\begin{minipage}[c]{.5\linewidth}
\includegraphics[width=1\textwidth]{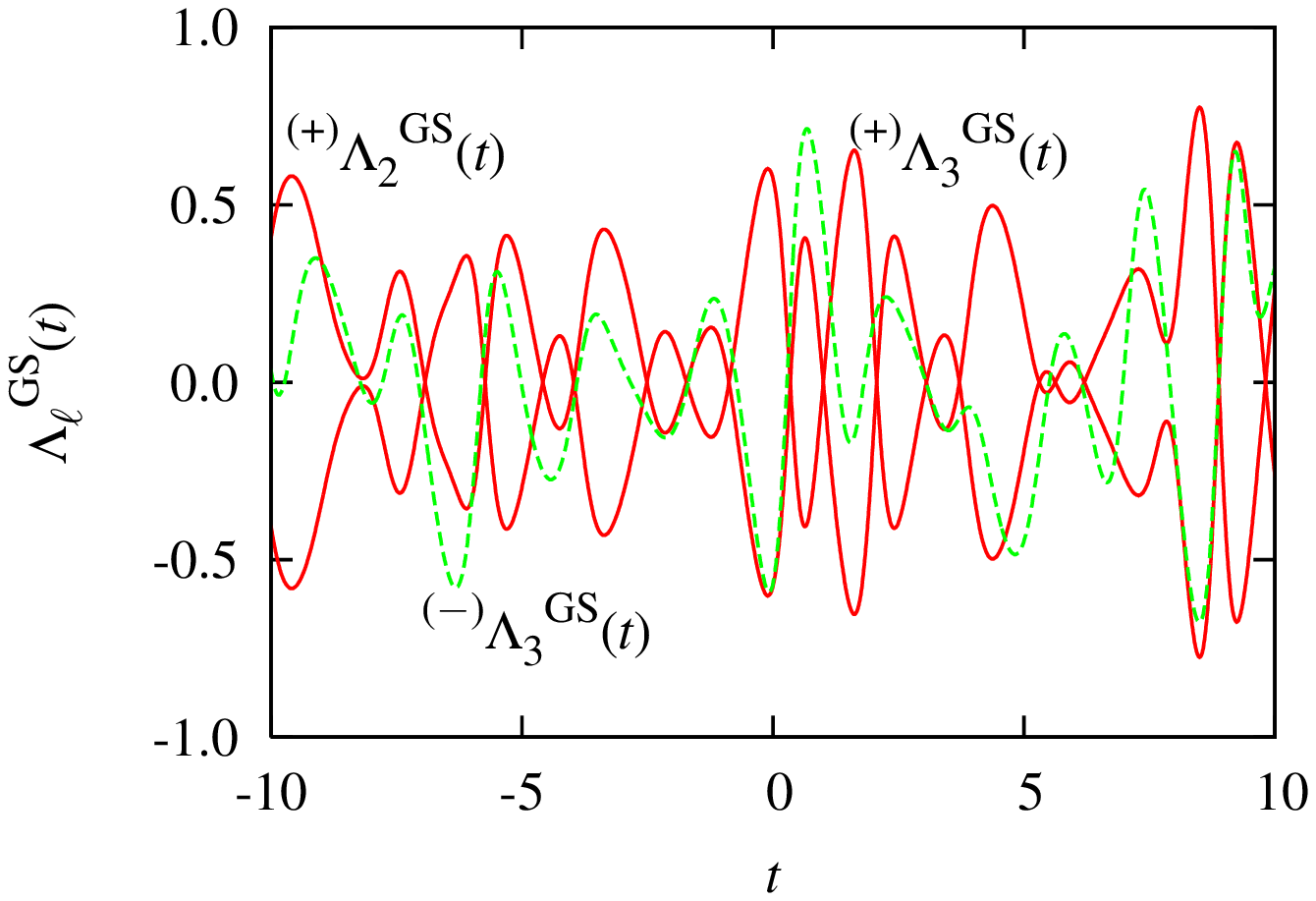}
\end{minipage} &
\begin{minipage}[c]{.5\linewidth}
\includegraphics[width=1\textwidth]{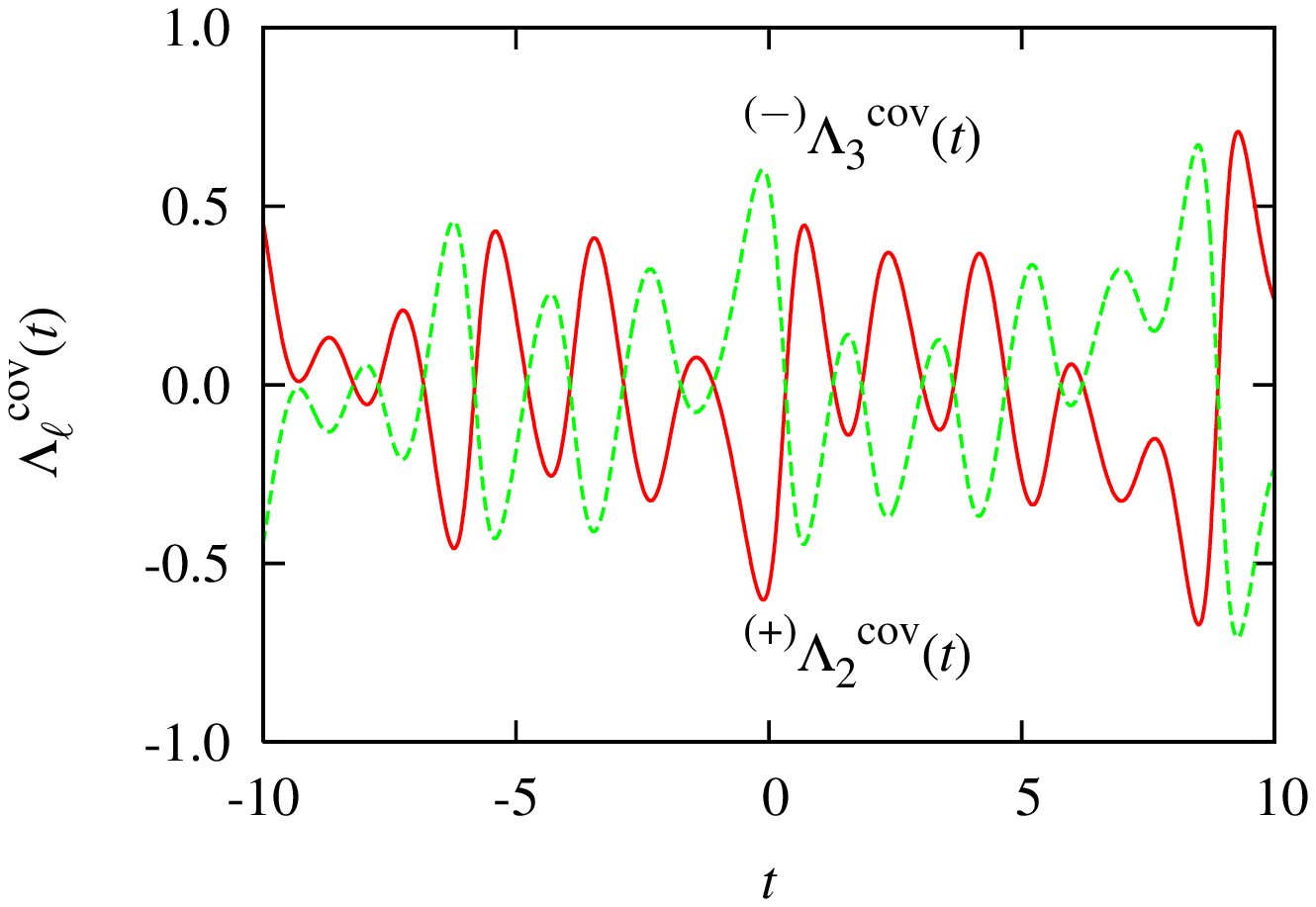}
\end{minipage} \\
\end{tabular}
\caption{(Color online) Local Lyapunov exponents for the H\'enon-Heiles system
with  energy $H = 1/6$. Left panel: Illustration of the symplectic local pairing symmetry,
Eq. (\ref{pair1}-\ref{pair2}), for the Gram-Schmidt (GS) exponents $^{(+)} \Lambda_2^{\mbox{GS}}$
and $^{(+)}\Lambda_3^{\mbox{GS}}$ (smooth red lines). The dimension of phase space  $D = 4$.
The inequality Eq. (\ref{inequ}) is demonstrated by the  dashed green
line for $^{(-)}\Lambda_3^{\mbox{GS}}$. Right panel: Verification of the 
time-reversal invariance property (\ref{local_symmetry_e}) for the local covariant exponents specified.} 
\label{figure_3}
\end{figure}
%%%%%%%%%%%%%%%%%%%%%%%%%%%%%%%%%%%%%%%%%%%%%%

In polar coordinates
defined by $ x = r \cos \phi $ and $y = r \sin \phi$ the Hamiltonian becomes
\begin{equation}
    H_P = \frac{p_r^2}{2} + \frac{p_{\phi}^2}{2 r^2} + \frac{r^2}{2} + \frac{r^3}{3} \sin 3\phi,
\end{equation}    
where, as before,  $p_r =  \dot{r} $ and $p_{\phi} =  r^2 \dot{\phi}$ are the conjugate momenta.
With an analysis completely analogous to that in the previous section for the spring pendulum,
the covariant vectors in the polar representation may be obtained. Here, it suffices to  present in
Fig. \ref{figure_4} only results for the polar
covariant vector $\bm{v}_P^1$ associated with the maximum exponent.
%%%%%%%%%%%%%%%%%%%%%%%%%%%%%%%%%%%%%%%%%%%%%%%%%%%
\begin{figure}[htb]
\begin{tabular}{c c}
\\
\begin{minipage}[c]{.5\linewidth}
\includegraphics[width=1\textwidth]{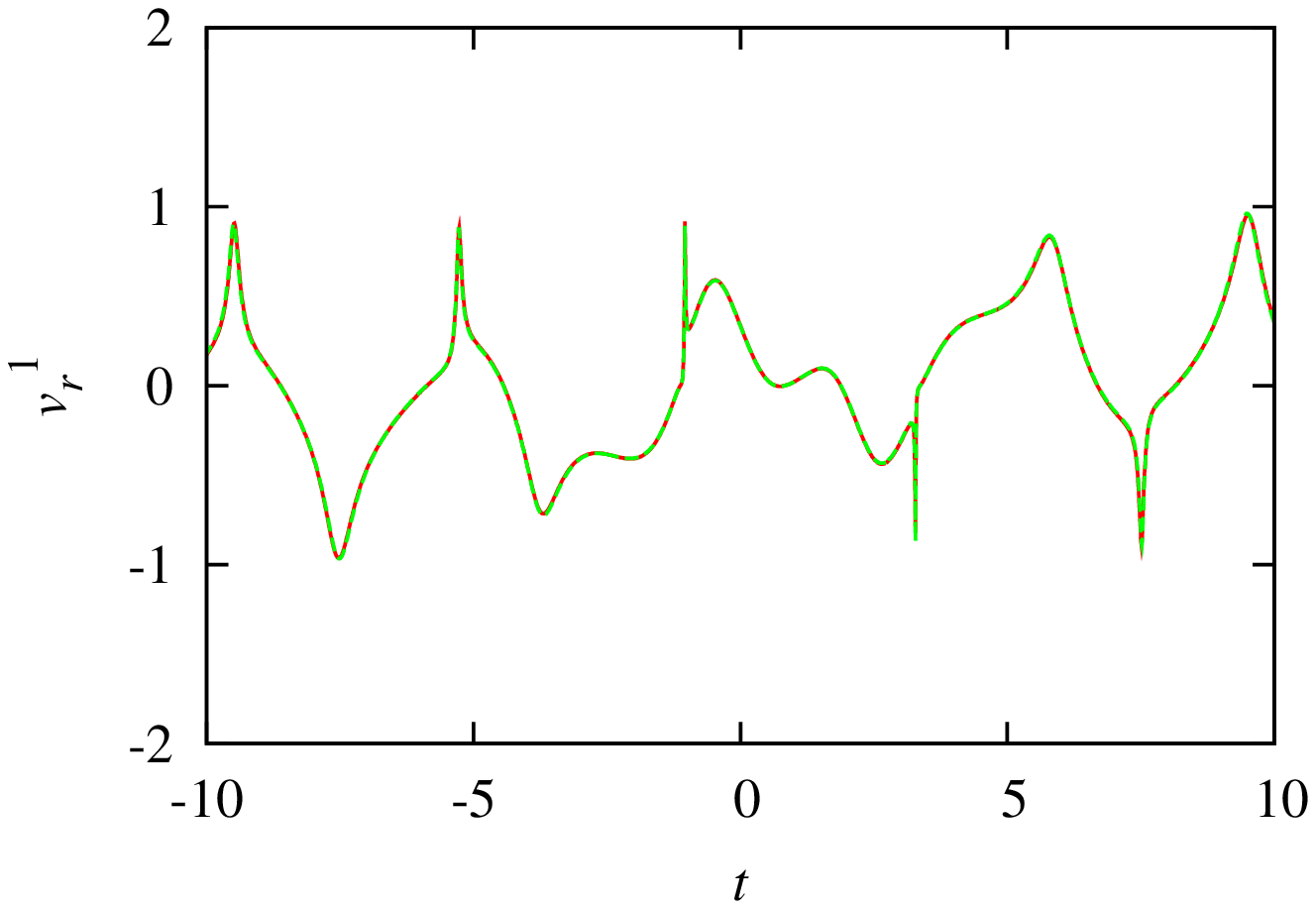}
\end{minipage} &
\begin{minipage}[c]{.5\linewidth}
\includegraphics[width=1\textwidth]{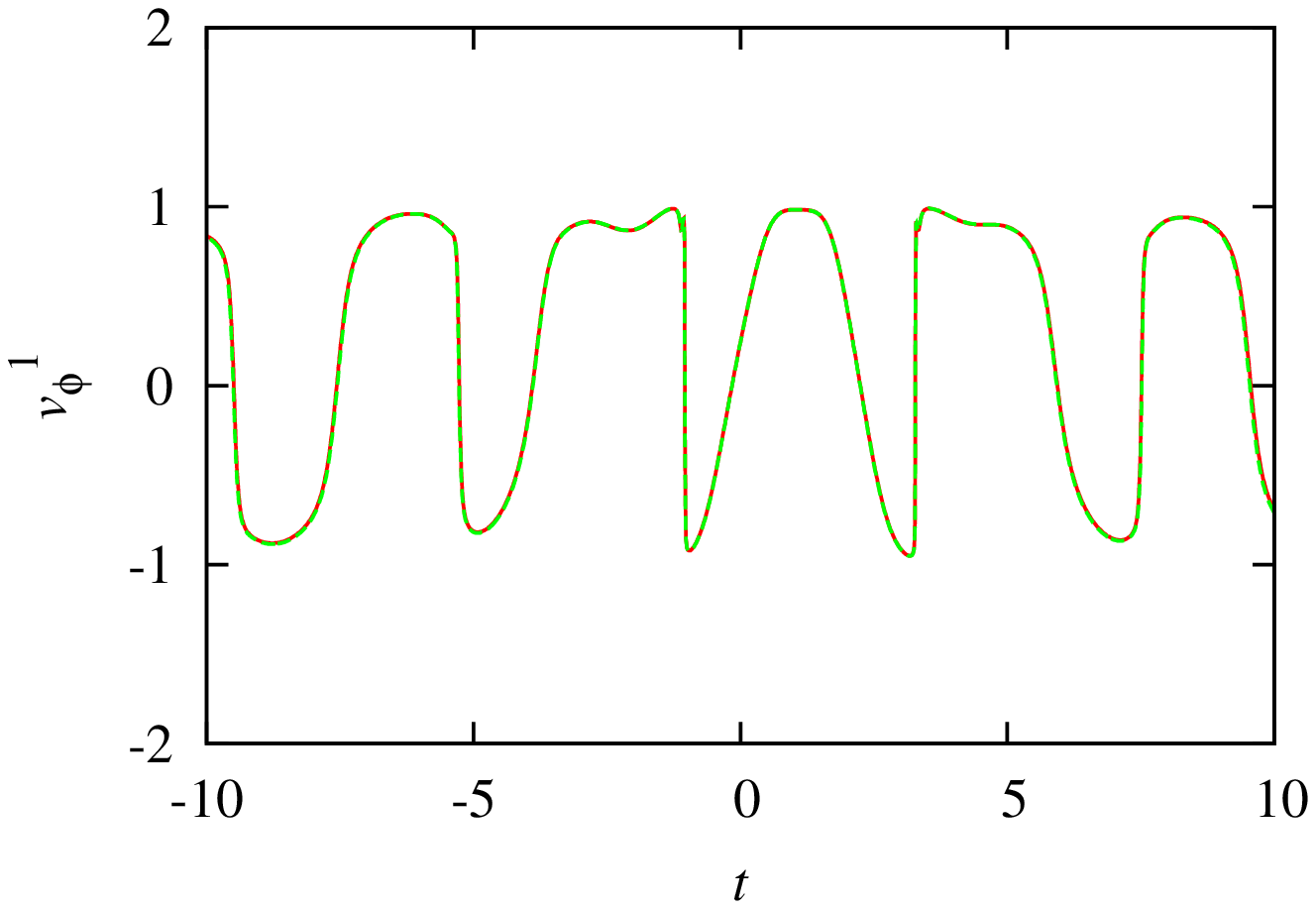}
\end{minipage} \\
\begin{minipage}[c]{.5\linewidth}
\includegraphics[width=1\textwidth]{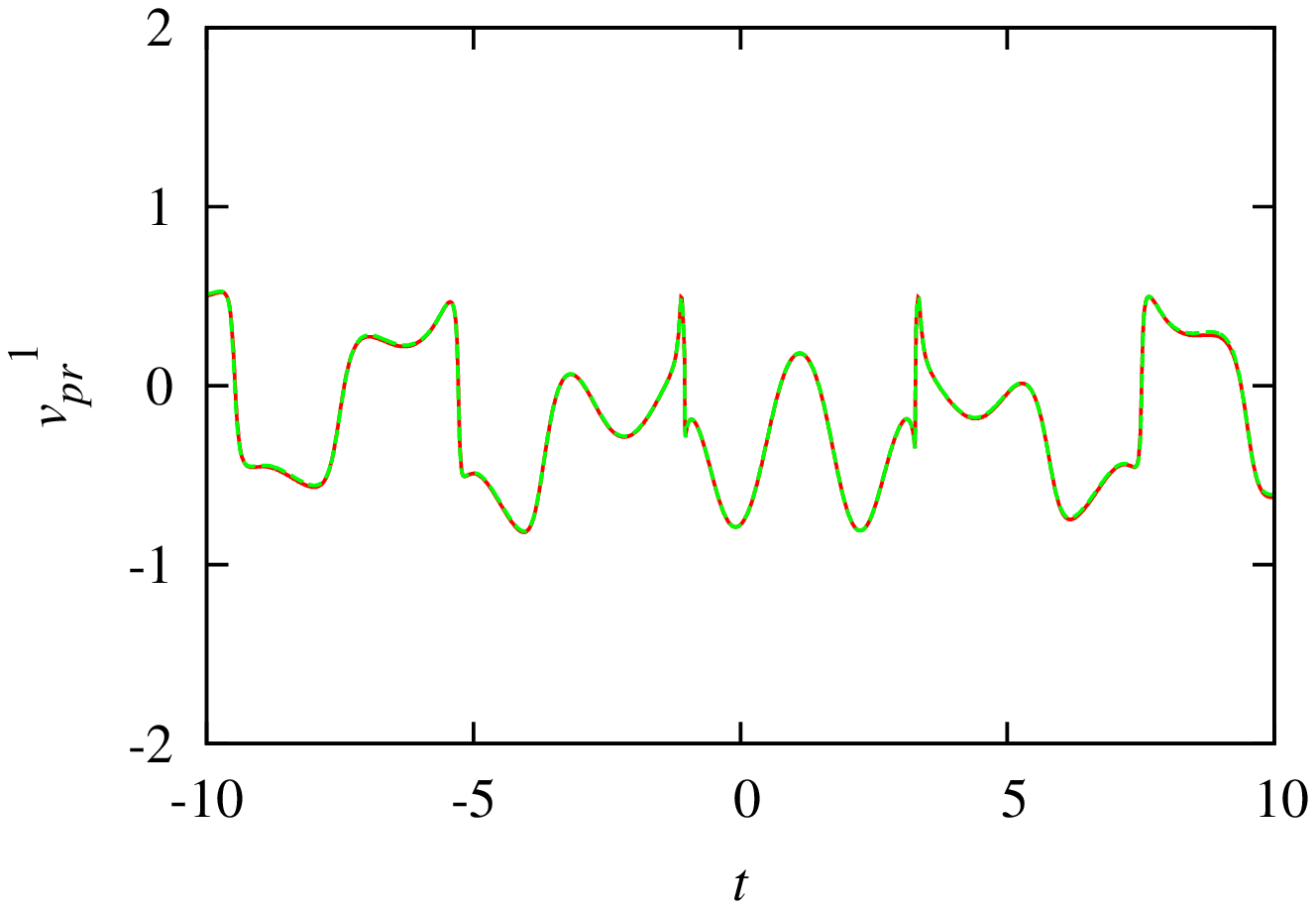}
\end{minipage}&
\begin{minipage}[c]{.5\linewidth}
\includegraphics[width=1\textwidth]{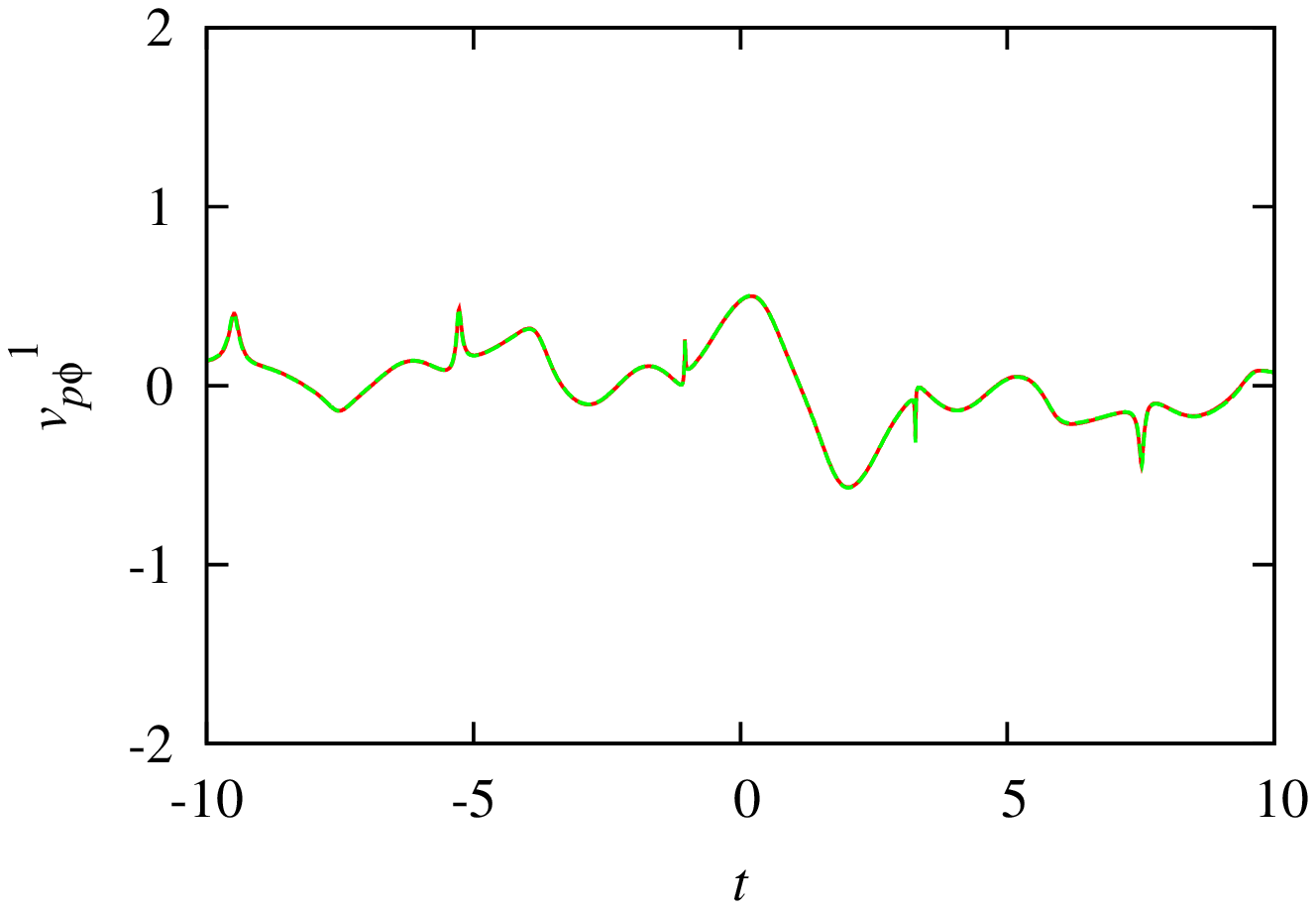}
\end{minipage}
\end{tabular}
\caption{(Color online) Time evolution of the 
four polar components $\delta r, \delta{\phi}, \delta p_r, \delta p_{\phi}$ (from top left to bottom right) for  the covariant vector  $\bm{v}_P^1$ spanning the unstable manifold of the H\'enon-Heiles system. 
In every panel two lines are shown. 
The smooth (red) lines are obtained by direct integration with polar coordinates, whereas the 
dashed (green) curves 
were converted from the Cartesian representation. Both lines agree and cannot be distinguished.}
\label{figure_4}
\end{figure}
%%%%%%%%%%%%%%%%%%%%%%%%%%%%%%%%%%%%%%%%%%%%%%%%%
One observes that the covariant vectors are unique (up to the sign as the example in the previous section testifies), and that it does not matter what 
parameterization is used for their computation.

\section{R\'esum\'e}

There are many recent applications of Lyapunov vectors, which range from weather forecasting \cite{Legras},
geophysical applications \cite{Wolfe}, and studies of transport in turbulent flow \cite{Fouxon1,Fouxon2},
to identifying and probing rare chaotic events by Lyapunov-weighted dynamics \cite{Kurchan}
and/or path sampling \cite{Geiger}.

From the two sets of perturbation vectors commonly used - the orthonormal Gram-Schmidt vectors and the co-moving covariant 
vectors - the latter are more attractive from a physics point of view. They reflect time-reversal invariance 
and, as is shown here, they are unique and independent of the  choice of the coordinate system.
By uniqueness we mean that they are properties of the flow in tangent space, and when they are 
known in a particular frame, they may be easily converted to another.
They provide a spanning set for the Oseledec splitting of the tangent space into a hierarchy of 
subspaces, each characterized by a well-defined Lyapunov exponent.
In this sense, they are a property of the physical system. This is not the case for the more artificial
ortho-normal Gram-Schmidt vectors, which are a mathematical device for computing 
volume changes of $d$-dimensional volume elements, $d \le D$, in the $D$-dimensional phase space.
Their advantage is that they are easier and faster to compute. According to the
algorithm of Ginelli {\em et al.} \cite{Ginelli}, the computation of the 
covariant vectors at a phase-point ${\bf \Gamma}$ requires the knowledge of the Gram-Schmidt 
vectors for all points on the trajectory through  ${\bf \Gamma}$ both in the past and the future \cite{Ruelle:1979}. 
Due to the availability of fast computers and programming techniques, rather  high-dimensional
systems have already been examined \cite{Yang_Radons_2010,rough,YRb}

For many problems only the  first Lyapunov vector associated with the maximum (global)
 exponent is required. Since the first Gram-Schmidt and the first covariant vectors always agree,
 one gets away with the fast computation of the first Gram-Schmidt vector, which is covariant.
 This advantage may dissolve again if the application requires the knowledge of the maximum
 {\em local} exponent together with its covariant vector. Due to possible exponent entanglement 
 \cite{YRa,BPDH} the maximum local exponent at a particular instant may belong, say,
 to the fifth vector. In such a case more than one covariant vectors need to be computed.    
\vspace{5mm}
\section{Acknowledgements}  The author  acknowledges stimulating discussions with 
 Hadrien Bosetti \cite{thesis} and William. G. Hoover and some very constructive remarks by
 an anonymous referee.
 
\end{document}